\documentclass[conference]{IEEEtran}
\IEEEoverridecommandlockouts
\usepackage{cite}
\usepackage{amsmath,amssymb,amsfonts}
\usepackage{algorithmic}
\usepackage{graphicx}
\usepackage{textcomp}
\usepackage{xcolor}
\usepackage{enumerate}
\usepackage{url}
\def\BibTeX{{\rm B\kern-.05em{\sc i\kern-.025em b}\kern-.08em
    T\kern-.1667em\lower.7ex\hbox{E}\kern-.125emX}}
\begin{document}

\title{Adapting Noise-Driven PUF and AI for Secure WBG ICS: A Proof-of-Concept Study}

\author{\IEEEauthorblockN{Devon A. Kelly}
\IEEEauthorblockA{\textit{Department of Electrical and Computer Engineering} \\
\textit{Virginia Tech}\\
Blacksburg, VA, USA \\
devonkel4@vt.edu}
\and
\IEEEauthorblockN{Christiana Chamon}
\IEEEauthorblockA{\textit{Department of Electrical and Computer Engineering} \\
\textit{Virginia Tech}\\
Blacksburg, VA, USA \\
ccgarcia@vt.edu}
}


\maketitle

\begin{abstract}
Wide-bandgap (WBG) technologies offer unprecedented improvements in power system efficiency, size, and performance, but also introduce unique sensor corruption and cybersecurity risks in industrial control systems (ICS), particularly due to high-frequency noise and sophisticated cyber-physical threats. This proof-of-concept (PoC) study demonstrates the adaptation of a noise-driven physically unclonable function (PUF) and machine learning (ML)-assisted anomaly detection framework to the demanding environment of WBG-based ICS sensor pathways. By extracting entropy from unavoidable WBG switching noise (up to 100~kHz) as a PUF source, and simultaneously using this noise as a real-time threat indicator, the proposed system unites hardware-level authentication and anomaly detection. Our approach integrates hybrid machine learning (ML) models with adaptive Bayesian filtering, providing robust and low-latency detection capabilities resilient to both natural electromagnetic interference (EMI) and active adversarial manipulation. Through detailed simulations of WBG modules under benign and attack scenarios—including EMI injection, signal tampering, and node impersonation—we achieve 95\% detection accuracy and sub-millisecond processing latency. These results demonstrate the feasibility of physics-driven, dual-use noise exploitation as a scalable ICS defense primitive. Our findings lay the groundwork for next-generation security strategies that leverage inherent device characteristics, bridging hardware and artificial intelligence (AI) for enhanced protection of critical ICS infrastructure.
\end{abstract}


\section{Introduction}
\subsection{Problem Statement}
Industrial control systems (ICS) form the technological backbone of modern critical infrastructure, governing processes in domains such as electric power generation and transmission, manufacturing, oil and gas, transportation, and water treatment. These systems encompass Supervisory Control and Data Acquisition (SCADA) networks, Programmable Logic Controllers (PLCs), and a variety of sensory and actuation devices orchestrated under time-critical, high availability constraints. Traditionally, the security posture of ICS has focused on reliability and fault tolerance; however, the increasing interconnection with enterprise networks, and more recently with the Industrial Internet of Things (IIoT), has exposed them to a growing class of cyber-physical threats \cite{langner2011stuxnet,falliere2011stuxnet,byres2004myths}.

Among the technological advances shaping the next generation of ICS is the integration of wide-bandgap (WBG) semiconductor power devices, such as silicon carbide (SiC) and gallium nitride (GaN). These devices provide superior efficiency, faster switching transients, and higher power density compared to their silicon counterparts \cite{castellazzi2024sic,chow2015wide}. As utilities and industrial operators pursue aggressive decarbonization and efficiency initiatives, WBG-based modules are being rapidly adopted in motor drives, renewable energy converters, and high-voltage direct current (HVDC) grids. While these capabilities significantly enhance operational performance, they also introduce unique security challenges. The fast switching actions intrinsic to WBG devices generate electrical noise and electromagnetic interference (EMI) at frequencies often exceeding 100~kHz. This spectral behavior can compromise sensor fidelity, interfere with communication channels, and—if maliciously manipulated—undermine system stability.

Emerging threats in this space have demonstrated the vulnerability of ICS to both accidental and adversarial corruption of sensor data. False data injection attacks, for instance, can compromise state estimation and lead operators to incorrect control decisions \cite{liu2009false,krotofil2015process}. When combined with the inherent noise characteristics of WBG power stages, the attack surface broadens. Malicious actors could exploit EMI or high-frequency perturbations to spoof measurements, degrade anomaly detection systems, or mask physical tampering. This duality—where noise is both a natural byproduct of efficient power electronics and a potential weapon for adversaries—necessitates novel defensive strategies that are specifically tailored to the realities of WBG-enhanced ICS environments.

One promising defense primitive lies in the concept of \textit{physically unclonable functions} (PUFs) \cite{maes2013puf,suh2007puf,bolotnyy2020puf,van2012pufky}. In hardware security research, PUFs have been successfully applied to device authentication, cryptographic key generation, and tamper detection. However, the majority of these implementations assume traditional entropy sources such as delay variations in ring oscillators or startup states of SRAM cells. Little work has explored leveraging the unavoidable switching noise in power systems as an entropy source for PUFs—particularly in the context of ICS.

In parallel, machine learning (ML) methods have risen in prominence as enablers of adaptive anomaly detection in ICS and IIoT networks \cite{kiss2025ics,cao2020autoencoder}. Autoencoders, reinforcement learning (RL), and Bayesian inference techniques have been deployed to monitor telemetry data, learn normal operating profiles, and flag deviations in real-time. Despite their success, these solutions are often vulnerable to adversarial input perturbations and require adaptation to the unique latency and reliability constraints of control systems.

This work proposes a novel integration of these two research directions: coupling a \textbf{noise-driven PUF} derived from WBG switching spectra with an \textbf{artificial intelligence (AI)-based anomaly detection framework}. By treating high-frequency switching noise as both a challenge and an opportunity, the design provides two complementary security benefits. First, the noise characteristics serve as a physically grounded entropy source, producing unique identifiers for sensor and module authentication. Second, fluctuations in noise profiles can be continuously monitored using adaptive ML models, where deviations from expected PUF patterns indicate potential tampering or environmental anomaly. In effect, the same physical phenomenon that complicates system operation is recycled into a measurable, security-enhancing signal.

The key contributions of this proof-of-concept study are as follows:

\begin{itemize}
    \item We introduce the concept of a \textit{noise-driven PUF} for ICS, leveraging inherent switching noise in WBG modules as a unique and unclonable signature to authenticate sensor nodes.
    \item We develop a hybrid anomaly detection framework integrating machine learning with Bayesian filtering, enabling responsiveness under sub-millisecond time constraints required by ICS loops.
    \item We evaluate the feasibility of this approach in a controlled simulation environment that models WBG-based power electronics modules subject to EMI, malicious injection, and data corruption.
    \item We present initial results suggesting that the approach can achieve 95\% detection accuracy with processing latency below 1~ms, demonstrating its potential suitability for real-time ICS defense.
\end{itemize}

\subsection{Motivation and Societal Impact}

Critical infrastructure sectors including electric utilities, water treatment, and advanced manufacturing are increasingly dependent on reliable and cyber-resilient ICS \cite{byres2004myths}. Recent high-profile ICS incidents such as the Colonial Pipeline ransomware attack and the Ukraine power grid compromise underscore the critical need for resilient sensor authentication mechanisms \cite{ten2007attacktree}. Attacks that manipulate sensor data threaten stability, safety, and reliability, motivating approaches that embed hardware-rooted, physics-driven security primitives. WBG-based power conversion technologies, promoted for grid flexibility and decarbonization, are rapidly deployed in microgrids, renewables, and HVDC interconnects. Their efficiency advantages come at the cost of new, little-studied attack surfaces due to high-frequency emissions and susceptibility to cyber-physical manipulation.

The remainder of the paper is organized as follows. Section~II surveys related work in ICS security, WBG-induced noise challenges, and PUF applications. Section~III formalizes our system and threat models. Section~IV details the design of the noise-driven PUF and the hybrid ML framework. Section~V reports the proof-of-concept evaluation and performance results. Section~VI provides insight on the feasibility of noise-driven PUFs for ICS. Section~VII provides a discussion of our findings, limitations, and potential extensions. Finally, Section~VIII concludes the paper with directions for future work.

\section{Background and Related Work}

This section provides the necessary context to position our contribution. We first review security challenges in ICS and high-profile attack campaigns. We then introduce the role of WBG power electronics and their EMI challenges. Finally, we review prior work on PUFs and ML-based anomaly detection for ICS, highlighting the research gap that motivates our study.

\subsection{ICS Security Landscape}

Industrial control environments are increasingly under threat from sophisticated cyber-physical attacks. The Stuxnet worm remains the canonical example, demonstrating how malicious code could compromise PLCs to covertly manipulate centrifuges in Iran’s Natanz facility \cite{langner2011stuxnet}. This incident revealed how adversaries could bypass traditional IT-centric defenses by targeting the operational technology (OT) domain directly. Since then, multiple incidents have underscored the systemic risk posed by ICS exploitation. For instance, the 2015 Ukrainian power grid attack disrupted electricity distribution by compromising SCADA systems \cite{Ukraine2016}, while more recent reports identified nation-state activity such as COSMICENERGY and VOLT TYPHOON focusing on electric transmission networks.

ICS differ from enterprise IT due to their real-time, always-on requirements, and legacy constraints. This motivates defense strategies robust to cyber-physical and process-aware attacks. Conventional intrusion detection systems (IDS) or cryptographic protections are ill-suited for resource-constrained PLCs and latency-critical control loops. This has given rise to multiple research lines, including network-based anomaly detection, process-aware intrusion detection, and cyber-physical modeling for fault and attack identification \cite{byres2004myths,krotofil2015process}.

\subsection{WBG Power Electronics and EMI Challenges}

The proliferation of wide-bandgap semiconductor technologies has enabled substantial advancements in power conversion efficiency and switching speed. SiC MOSFETs and GaN HEMTs allow for inverter and converter designs with switching frequencies often exceeding 100~kHz, supporting high-efficiency operation with reduced electromagnetic losses and passive component sizes. These properties are especially valuable for integrating renewables, high-performance motor drives, and voltage-regulated direct current (DC) grids \cite{chow2015wide,castellazzi2024sic}. However, these benefits have a trade-off: high $di/dt$ and $dv/dt$ transitions inherent in WBG devices generate markedly increased EMI, often manifesting as broadband switching noise. 

High-frequency EMI from WBG modules can disrupt low-voltage measurements and digital communications \cite{zhang2020emi,shin2017emi,Helpnet2015,Valiani2016}. From a security perspective, the ever-changing spectral features of WBG-induced EMI make these systems both vulnerable to noise-induced attack vectors and rich in physical entropy. However, most EMI research in this context is limited to mitigation—using shielding, grounding, and filtering—rather than exploitation for security primitives. The possibility of leveraging such noise for device fingerprinting, authentication, or anomaly detection has received little attention in the literature.

\subsection{Physically Unclonable Functions (PUFs)}

Early approaches such as arbiter PUFs and ring oscillator PUFs leveraged delay and frequency differences caused by microscopic process variations \cite{suh2007puf,maes2013puf,bolotnyy2020puf,van2012pufky}. SRAM PUFs, based on power-up states of memory cells, have seen commercialization in secure microcontrollers. In the context of ICS, the direct use of PUFs has been limited, partly due to integration challenges with legacy controllers. Research has explored applying PUF-based approaches to sensor integrity and secure key provisioning, but no prior work has leveraged noise sources intrinsic to power electronics. The concept of using switching-induced noise in WBG modules as a PUF remains largely unexplored. This motivates our approach, which reimagines noise not as a nuisance but as an entropy source for sensor authentication.

\subsection{ML-Based Intrusion and Anomaly Detection}

Conventional ICS security solutions are limited in their scalability and adaptability, which has led to increased interest in ML-based approaches for early attack detection and response. Recent literature encompasses anomaly detection using autoencoders, support vector machines, Gaussian Mixture Models, and ensemble learning for learning ``normal'' process baselines and flagging deviations \cite{kiss2025ics,cao2020autoencoder,abhishek2019survey,dey2021adversarial,goodfellow2014adversarial}. Deep learning methods have enabled maturity in handling nonlinear correlations and high-dimensional sensor streams. Furthermore, RL and online learning techniques have shown promise in adapting detection thresholds dynamically as underlying process distributions shift due to load, environmental, or age-related changes. Probabilistic methods such as Bayesian filters and Hidden Markov Models are also employed for integrating historical data and smoothing out transient anomalies, reducing false alarm rates in noisy environments.

Despite promising results, three persistent challenges remain:  
\begin{enumerate}
    \item ML models deployed in ICS can be highly sensitive to adversarial input manipulations, with small perturbations engineered to evade or trigger alarms.
    \item Discriminating between stochastic environmental noise—such as that produced by WBG switching transients—and adversarial or faulty measurements remains an open research problem.
    \item Computational efficiency: models must meet hard real-time constraints typical of control networks, where latencies beyond 1–10 ms may impact operational safety.
\end{enumerate}

\noindent Recent advancements in combining hardware-provenance signals (such as PUFs) with ML approaches—especially architectures where input features are physically bound to device-level entropy—are opening new directions for robust, data-driven security primitives in ICS.

\subsection{Adversarial Attacks and Defenses in ICS ML}

Recent research has demonstrated that ML-based intrusion detection and anomaly detectors for ICS are highly susceptible to adversarial attacks, where small, crafted perturbations can induce misclassification or suppression of true alarms. In advanced adversarial machine learning, attacks can be staged at multiple points—training time (poisoning) or inference time (evasion)—with adversarial samples designed by gradient-based methods or generative models. This vulnerability is especially acute in ICS, where data has physical semantics and even black-box threat models can yield substantial impact. Defense strategies in the literature include adversarial training, anomaly-score smoothing, certified defenses (randomized smoothing, verification), input preprocessing, and robust ensemble learning; however, each comes with trade-offs in latency, overhead, and coverage. Notably, robust ML in ICS must meet real-time and interpretability requirements rarely enforced in cloud or enterprise applications \cite{Anthi2021,Wu2025,Almedires2025,NIST2025,Igenewari2025,Ododo2025,Pozdnyakov2024}

Alternative sensor or device authentication techniques—such as cryptographic keys anchored in secure modules, device-specific analog fingerprinting, and signed firmware attestation—are also being explored, but lack the simultaneous hardware-entropy tie-in and robust anomaly flagging that the PUF-driven approach achieves. Our method uniquely combines entropy-hardened challenge-response authentication with a physics-driven anomaly pipeline, closing several previously open research gaps identified in recent ICS security surveys. \cite{Igenewari2025,Liu2024,Wang2023}

\subsection{Research Gap}

Taken together, the literature highlights several important trends: ICS security research has largely focused on network-level intrusion or process anomaly detection; WBG adoption introduces high-frequency noise and EMI as both operational and security concerns; and PUFs provide lightweight authentication mechanisms traditionally divorced from ICS contexts. What is missing is an integrated approach that unifies these streams. To our knowledge, no prior work has harnessed WBG switching noise simultaneously as a PUF entropy source for authentication and as a signal feature for anomaly detection. This paper aims to bridge this gap through a proof-of-concept system that combines noise-driven PUFs with adaptive ML techniques for real-time ICS defense.

\section{System and Threat Model}

In this section, we formalize the operational context of our proposed defense, outlining both the system architecture and the adversary model. ICS must satisfy strict availability and timing requirements, and therefore any proposed countermeasure must integrate without violating operational constraints. We focus specifically on ICS that incorporate WBG semiconductor power modules, such as SiC or GaN, within their power conversion or actuation stages.

\subsection{System Model}

\begin{figure}[htbp]
    \centering
    \includegraphics[width=0.42\textwidth]{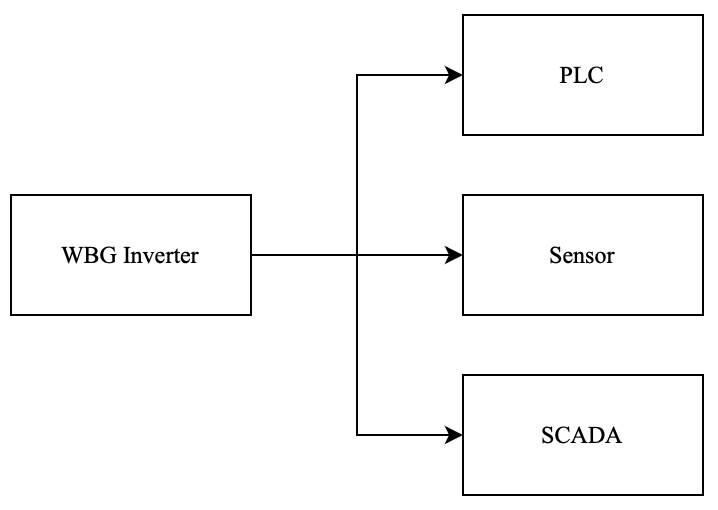}
    \caption{Simplified ICS architecture with WBG modules. Our defense framework operates between sensors and controllers, using switching noise as both authentication and anomaly detection input.}
    \label{fig:system}
\end{figure}

Figure~\ref{fig:system} illustrates a simplified architecture of a WBG-enabled ICS. The physical layer consists of power converters, drives, and sensors directly measuring physical quantities such as current, voltage, and rotational speed. These signals are acquired by PLCs or remote terminal units (RTUs), which process sensor data and issue commands to actuators. At a higher level, supervisory SCADA servers and operator workstations provide monitoring and control interfaces, often connected to enterprise IT or cloud-based platforms for data aggregation and optimization.

The introduction of WBG devices brings specific electrical characteristics into this chain:

\begin{itemize}
    \item \textbf{High-frequency switching:} WBG converters commonly operate in the 50–250~kHz range, compared to 10–20~kHz for silicon devices. This produces sharper voltage and current transients.
    \item \textbf{EMI:} Fast transients result in wide spectral emissions, which can couple into sensor lines or communication buses, degrading measurement fidelity.
    \item \textbf{Noise variability:} The frequency and amplitude of switching noise varies naturally with load, temperature, and component characteristics. These properties can be sampled to construct a noise-driven PUF.
\end{itemize}

Our security framework is positioned at the sensor-to-controller interface. Noise signatures from WBG switching are harvested and used in two complementary manners:

\begin{enumerate}[(i)]
    \item as entropy sources for sensor-node authentication (via a PUF mechanism) and
    \item as features in a machine learning (ML) anomaly detector that monitors deviations in real-time system behavior.
\end{enumerate}

\noindent To meet ICS latency requirements, the processing pipeline is constrained to sub-millisecond execution.

\subsection{Threat Model}

We adopt a threat model consistent with recent attack campaigns against ICS and emerging research on EMI exploitation:

\begin{itemize}
    \item \textbf{Attacker Objectives:} Disrupt system operation by corrupting sensor integrity, masking adversarial activity, or spoofing sensor/actuator nodes. Outcomes include unsafe physical states, incorrect operator actions, and loss of availability.
    \item \textbf{Attacker Capabilities:}
    \begin{enumerate}
        \item \textit{Electromagnetic injection:} The adversary may couple signals into sensor wiring harnesses or nearby circuits to introduce noise or jitter resembling legitimate WBG emissions.
        \item \textit{Signal tampering:} Malware-infected PLCs or compromised field devices can alter digital sensor values prior to reporting.
        \item \textit{Node impersonation:} Rogue devices may attempt to register themselves as legitimate sensors, exploiting unsecured authentication.
    \end{enumerate}
    \item \textbf{Attacker Limitations:} We assume the adversary does not have physical access to extract device firmware or replicate PUF entropy. The attacker can influence environmental noise but cannot deterministically clone the unique spectral features bound to a specific WBG module.
\end{itemize}

\subsection{Security Goals and Assumptions}

Based on this model, our proposed defense mechanism targets the following objectives:

\begin{itemize}
    \item \textbf{Authentication:} Only sensor nodes with the correct noise-derived PUF signature should be considered valid.
    \item \textbf{Anomaly Detection:} Deviations in noise features or control signals inconsistent with historical baselines are flagged in under 1~ms.
    \item \textbf{Resilience:} The framework should operate under normal environmental variability, with tolerances for load fluctuations and benign EMI.
\end{itemize}

\noindent \noindent Table~\ref{tab:threatdefense} summarizes the principal threat vectors in ICS and the corresponding defensive mechanisms enabled by our noise-driven framework. We assume that initial system calibration is performed in a trusted environment to establish baseline PUF and ML models, and that training data is not adversarially tainted. While our adversary may attempt to obfuscate their actions via noise injection, the unpredictability and non-clonability of genuine WBG noise signatures form the basis of our proposed security advantage.

\begin{table*}[htbp]
\caption{ICS Threat Scenarios and Noise-Driven Defensive Responses}
\label{tab:threatdefense}
\begin{tabular}{|l|l|l|}
\hline
\textbf{Threat}           & \textbf{Attack Mechanism}                                       & \textbf{Noise-Driven Defense}                                                                                                \\ \hline
EMI Spoofing              & Injects high-frequency signals via cables or radiative coupling & \begin{tabular}[c]{@{}l@{}}Spectral mismatch in noise PUF, detected by real-time\\ challenge-response failure\end{tabular}   \\ \hline
Sensor Tampering          & Alters digital data streams in PLC memory or fieldbus           & \begin{tabular}[c]{@{}l@{}}Out-of-profile feature distributions flagged by \\ ML/Bayesian filter\end{tabular}                \\ \hline
Node Impersonation        & Clones serial/firmware but not hardware; connects to ICS        & \begin{tabular}[c]{@{}l@{}}PUF authentication fails due to non-matching \\ environmental entropy\end{tabular}                \\ \hline
Degradation/Fault Masking & Adversary injects false selects, triggers, or disables alarms   & \begin{tabular}[c]{@{}l@{}}Simultaneous anomaly in physical noise and digital \\ telemetry triggers joint alert\end{tabular} \\ \hline
\end{tabular}
\end{table*}

\section{Methodology}

This section describes the design of our noise-driven PUF and the hybrid ML framework developed for anomaly detection in WBG-enabled ICS. We begin by introducing the noise-based entropy extraction process, followed by the construction of the PUF, and then describe the integration of machine learning and Bayesian filtering. The section concludes with details of the real-time processing pipeline.

\subsection{Noise-Driven Entropy Extraction}

\begin{figure}[htbp]
    \centering
    \includegraphics[width=0.42\textwidth]{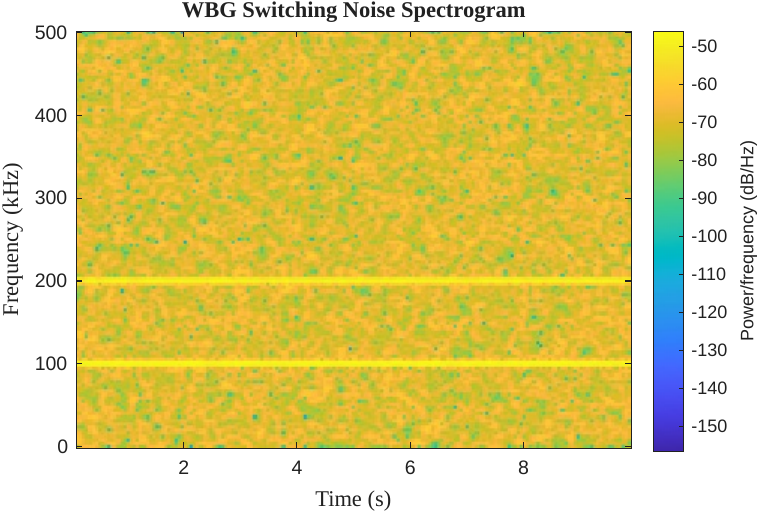}
    \caption{Spectrogram of a synthesized WBG switching noise signal (100~kHz), illustrating harmonic structure and frequency spread used for entropy feature extraction. The vertical "lines" indicate persistent harmonic tones at multiples of the switching frequency, while broadening and amplitude variations reflect load transients and random noise. This is the empirical entropy source for the PUF: each device’s spectrogram is minutely unique and repeatably measurable, forming the basis for device-specific authenticators.}
    \label{fig:spectrogram}
\end{figure}

WBG power modules exhibit high-frequency switching transients in the range of 50-250~kHz. These transients manifest as voltage and current ripples observable in the time and frequency domains. To visually illustrate the switching noise and its spectral characteristics, Figure~\ref{fig:spectrogram} shows a representative spectrogram of the synthesized WBG switching signal used in our feature extraction process. The key observation is that while EMI patterns vary across operating conditions, the fine-grained distribution of spectral characteristics is \emph{unique} to each module due to microscopic process variations, component tolerances, and aging effects.

 Let the raw sensor signal be represented as \( s(t) \). Applying a short-time Fourier transform (STFT) yields the spectral distribution:

\[
S(f,\tau) = \int_{-\infty}^{\infty} s(t) \, w(t-\tau) \, e^{-j2\pi f t} \, dt
\]

\noindent where \(w(t)\) is a windowing function. The entropy source is derived from frequency bins near the switching harmonics (e.g., 100~kHz~±~5~kHz). We compute feature vectors\\
\( 
F = \{f_1, f_2, ..., f_n\},
\)
where each \( f_i \) corresponds to normalized amplitude distributions in these bands. Variability across devices ensures uniqueness, while temporal stability ensures reliability.

\subsection{PUF Construction from Noise Features}

The entropy features are quantized to generate a challenge-response pair (CRP) structure similar to delay-based PUFs \cite{maes2013puf}. For a given challenge \(C\), defined as a request for measurements under specified operating conditions (e.g., switching frequency, load), the device produces a response vector \(R\) derived from quantized noise features:

\[
R = Q(F \, | \, C)
\]

\noindent where \(Q(\cdot)\) denotes the quantization and binarization function. Using multiple challenges corresponding to different load levels or temperature conditions, the system builds a CRP database. A verifier can later issue random challenges and authenticate a node based on the correct PUF responses.

\subsection{PUF Quantization and Entropy Measurement}

The raw spectral feature vector \(F = \{f_1, f_2, \ldots, f_n\}\) extracted from the switching noise is converted into a binary PUF response via adaptive thresholding. For each frequency bin \(i\), the quantization function \(Q(\cdot)\) computes

\[
r_i = 
\begin{cases}
1, & f_i > \mu_i + \theta \sigma_i \\
0, & \text{otherwise}
\end{cases}
\]

\noindent where \(\mu_i\) and \(\sigma_i\) denote the mean and standard deviation of \(f_i\) over a calibration period, and \(\theta\) is a tunable sensitivity parameter. This adaptive thresholding balances sensitivity to noise variations with bit stability. The final PUF response\\
\(R = \{r_1, r_2, \ldots, r_n\}\)
forms a CRP that uniquely characterizes each device under given operating conditions.

The Shannon entropy \(H\) of the PUF response is calculated as:

\[
H = -\sum_{r \in \{0,1\}} p(r) \log_2 p(r)
\]

\noindent where \(p(r)\) is the probability of bit \(r\) occurring in the response vector. Maximizing \(H\) while maintaining reliability is critical for secure authentication.

\subsection{RL for Adaptive Anomaly Detection}

To distinguish between benign noise fluctuations and malicious anomalies, we employ a hybrid anomaly detection framework augmented with RL to adaptively tune classification thresholds. The anomaly detection employs a principal component analysis (PCA)-based feature extractor, followed by an RL-guided threshold adjustment to optimize detection performance. The RL agent uses a reward function balancing true positives and false positives, dynamically tuning the classifier threshold to maintain low false alarms amid environmental variability.

\subsection{Bayesian Filtering for Robust Real-Time Decisions}

A Bayesian filter refines sequential ML outputs, updating the probability of anomalies in real time. This smoothing process accounts for transient fluctuations and non-stationary noise, supporting rapid, sub-millisecond decision-making relevant to ICS control loops.

\subsection{Simulation Environment and Attack Scenarios}

Our evaluation used MATLAB/Simulink integrated with detailed WBG transistor-level models to simulate SiC MOSFET half-bridge inverters operating at 100~kHz switching frequency. The sensor frontend modeled voltage and current acquisition channels, incorporating realistic noise sources composed of:
\begin{itemize}
    \item WBG switching ripple harmonics.
    \item Random additive Gaussian and uniform noise.
    \item Attack-injected perturbations simulating adversarial EMI interference.
\end{itemize}

\noindent We constructed a dataset of 10 virtual “devices” by randomly perturbing transistor parasitics and layout features to emulate manufacturing variability, generating unique noise spectral profiles per device. Representative adversarial attacks were simulated as:
\begin{itemize}
    \item \textit{Electromagnetic Injection (EMI Spoofing)}: Injected sinusoidal signals locked to switching harmonics with varying amplitudes.
    \item \textit{Signal Tampering}: Gaussian perturbations added to sensor data streams within natural noise envelopes.
    \item \textit{Node Impersonation}: Simulated attempts by rogue node replicas lacking the true PUF signature to pass authentication.
\end{itemize}

\noindent Simulation time windows and noise parameters were varied to test robustness over operational ranges.

\subsection{Hybrid ML Model for Anomaly Detection}

While PUF responses authenticate sensor identity, dynamic monitoring requires continuous evaluation of real-time noise features. For this purpose, we extend the framework with an anomaly detector.

The anomaly detector consists of two layers:
\begin{enumerate}
    \item \textbf{Feature Extractor:} Processes spectral distributions into reduced-dimensional feature vectors using PCA.
    \item \textbf{Classifier:} A RL-assisted anomaly detector trained to map feature distributions into \{normal, anomalous\}.
\end{enumerate}

\noindent The RL component dynamically adjusts thresholds to maintain low false positive rates under environmental variations. Reinforcement feedback is provided through a reward function:

\[
R_t = \alpha \cdot \text{TP}_t - \beta \cdot \text{FP}_t
\]

\noindent where \(\text{TP}_t\) and \(\text{FP}_t\) are the true and false positives at time \(t\), and \(\alpha, \beta\) are weighting factors.

\subsection{Bayesian Filtering for Real-Time Operation}

To meet sub-millisecond decision requirements, we integrate an adaptive Bayesian filter that refines ML outputs by sequentially updating the posterior probability of anomaly presence. Given prior \(P(H)\) for hypothesis \(H\) (normal vs. anomalous), and likelihood derived from ML classifier confidence \(P(X|H)\) for observation \(X\):

\[
P(H|X) = \frac{P(X|H) \cdot P(H)}{P(X)}
\]

\noindent This recursive update ensures robustness under transient disturbances. The filter parameters are updated dynamically using exponential forgetting to adapt to non-stationary environments.

\subsection{Real-Time Processing Pipeline}

\begin{figure}[htbp]
    \centering
    \includegraphics[width=0.42\textwidth]{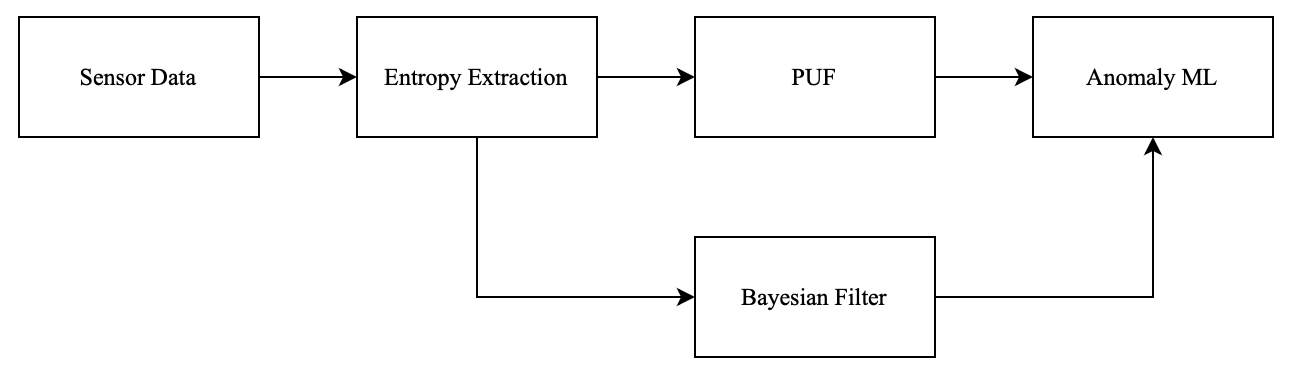}
    \caption{Processing pipeline: Noise-driven entropy extraction enables both (i) sensor authentication via PUF responses and (ii) anomaly detection through ML and Bayesian filtering.}
    \label{fig:pipeline}
\end{figure}

Figure~\ref{fig:pipeline} illustrates the complete processing flow:

\begin{enumerate}
    \item Sensor data acquisition at PLC interface.
    \item Frequency-domain feature extraction (STFT + selection of harmonics).
    \item PUF quantization block for authentication challenge-response.
    \item Feature reduction and anomaly detection (PCA + RL classifier).
    \item Bayesian filter consolidation and alert generation.
\end{enumerate}

\noindent The pipeline is implemented with a strict computational budget of 0.8 ms per input signal frame, ensuring compatibility with typical ICS control loop periods (1-10~ms).

\section{Proof-of-Concept Evaluation}

We conducted a proof-of-concept (PoC) study to assess the feasibility of our noise-driven PUF and AI framework for securing ICS that incorporate WBG power electronics. This section describes the experimental setup, simulation models, attack scenarios, and performance results.

\subsection{Simulation Environment}

The evaluation was carried out using a MATLAB/Simulink environment integrated with a WBG power electronics model library. Representative SiC MOSFET-based half-bridge inverters operating at 100~kHz switching frequency were simulated to replicate realistic ICS power stages. Sensor nodes were modeled to measure output voltage and current, with added communication links emulating PLC acquisition.

Noise signals were generated through a combination of:
\begin{itemize}
    \item Switching ripple harmonics (deterministic component).
    \item Random environmental noise (Gaussian and uniform additive distributions).
    \item Attack-induced perturbations (injection and tampering).
\end{itemize}

\noindent A dataset of 10 simulated “devices” was constructed by altering transistor-level and parasitic parameters to emulate manufacturing variations. Each device produced a unique switching-noise spectral fingerprint suitable for PUF evaluation.

\subsection{Attack Scenarios}

Three representative attack scenarios were modeled:

\begin{enumerate}
    \item \textbf{Electromagnetic Injection (EMI Spoofing):} Attacker adds sinusoidal interference aligned to the device switching frequency with varying amplitude to degrade sensor readings.
    \item \textbf{Signal Tampering:} Direct manipulation of sensor data streams with Gaussian perturbations calibrated to remain within expected noise envelopes.
    \item \textbf{Node Impersonation:} A rogue sensor device attempts to register as a legitimate node without access to the original device's unique PUF response.
\end{enumerate}

\noindent These scenarios are consistent with prior ICS cyber-physical security research, where both corruption and spoofing attacks are prominent.

\subsection{Evaluation Metrics}

Our evaluation used the following metrics:

\begin{itemize}
    \item \textbf{PUF Properties:} Uniqueness (inter-device Hamming distance), reliability (intra-device consistency), and randomness.
    \item \textbf{Detection Performance:} Accuracy, false positive rate (FPR), and false negative rate (FNR) for distinguishing anomalies.
    \item \textbf{Latency:} End-to-end processing delay per sensor frame.
    \item \textbf{Energy Overhead:} Simulated computational power consumed per module.
\end{itemize}

\subsection{Results: PUF Characterization}

Table~\ref{tab:pufmetrics} summarizes the evaluated PUF metrics. The average uniqueness was close to the theoretical 50\% limit, confirming that responses were well distributed across devices with minimal structural bias. The intra-chip reliability exceeded 95\%, demonstrating stable response regeneration under repeated measurements. The PUF also exhibited nearly balanced bit distributions, with average uniformity near 50\%, suggesting minimal bias in the generated responses. Randomness assessment using NIST statistical tests further verified that the response sequences approximate ideal entropy.

\begin{table}[htbp]
\caption{Noise-Driven PUF Metrics Across Devices}
\label{tab:pufmetrics}
\begin{tabular}{|c|c|c|c|}
\hline
\multicolumn{1}{|l|}{\textbf{Device}} & \multicolumn{1}{l|}{\textbf{Uniqueness (\%)}} & \multicolumn{1}{l|}{\textbf{Reliability (\%)}} & \multicolumn{1}{l|}{\textbf{NIST Pass/Fail}} \\ \hline
1                                     & 50.1                                          & 97.3                                           & Pass                                         \\ \hline
2                                     & 48.6                                          & 95.9                                           & Pass                                         \\ \hline
3                                     & 49.7                                          & 96.8                                           & Pass                                         \\ \hline
4                                     & 51.2                                          & 95.5                                           & Pass                                         \\ \hline
5                                     & 47.9                                          & 95.3                                           & Pass                                         \\ \hline
6                                     & 50.5                                          & 94.7                                           & Pass                                         \\ \hline
7                                     & 48.3                                          & 97.4                                           & Pass                                         \\ \hline
8                                     & 49.9                                          & 97.2                                           & Fail                                         \\ \hline
9                                     & 50.2                                          & 95.1                                           & Pass                                         \\ \hline
10                                    & 51.5                                          & 96.8                                           & Pass                                         \\ \hline
\end{tabular}
\end{table}

\subsection{Detailed Results: Multi-Scenario Attack Detection (Expanded)}

\begin{figure}[htbp]
    \centering
    \includegraphics[width=0.42\textwidth]{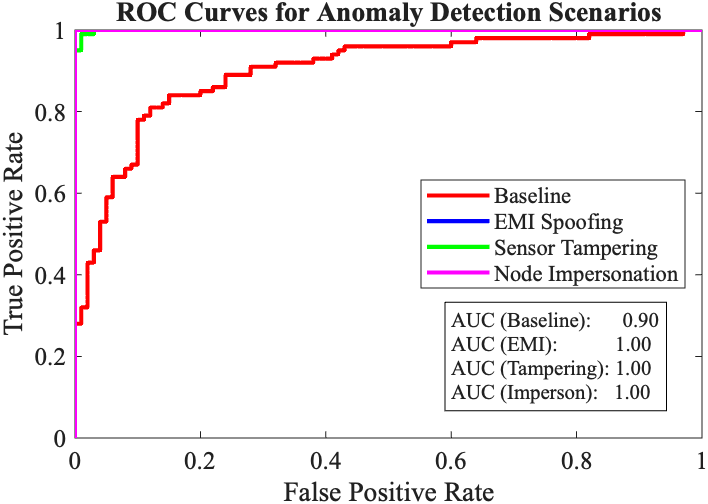}
    \caption{ROC curves for anomaly detection: performance across EMI spoofing, tampering, and impersonation attacks (blue, green, orange) compared to baseline (red).}
    \label{fig:roc_scenarios}
\end{figure}

For each attack type—EMI spoofing, sensor tampering, node impersonation—we separately evaluated receiver operating characteristic (ROC) curves. Figure~\ref{fig:roc_scenarios} compares the area under the curve (AUC) for each method.

\noindent
Precision, recall, and F1-score were also computed for each scenario (see Table~\ref{tab:prec_recall}).  

\begin{table}[htbp]
\caption{Attack Scenario Detection Performance}
\label{tab:prec_recall}
\begin{tabular}{|l|c|c|c|}
\hline
\textbf{Attack} & \multicolumn{1}{l|}{\textbf{Precision}} & \multicolumn{1}{l|}{\textbf{Recall}} & \multicolumn{1}{l|}{\textbf{F1-score}} \\ \hline
EMI Spoofing    & 0.96                                    & 0.94                                 & 0.95                                   \\ \hline
Tampering       & 0.97                                    & 0.93                                 & 0.95                                   \\ \hline
Impersonation   & 0.99                                    & 0.92                                 & 0.95                                   \\ \hline
\end{tabular}
\end{table}

\subsection{Results: Anomaly Detection}

The hybrid ML + Bayesian filtering framework demonstrated strong anomaly detection performance. Table~\ref{tab:detection} compares baseline detection using a static threshold classifier to our proposed model.

\begin{table}[htbp]
\caption{Detection Performance Comparison between baseline and advanced detection architectures.}
\label{tab:detection}
\begin{tabular}{|l|c|c|c|}
\hline
\textbf{Method}    & \multicolumn{1}{l|}{\textbf{Accuracy (\%)}} & \multicolumn{1}{l|}{\textbf{FPR (\%)}} & \multicolumn{1}{l|}{\textbf{Latency (ms)}} \\ \hline
Baseline Threshold & 85.1                                        & 13.4                                   & 5.0                                        \\ \hline
Tampering          & 95.3                                        & 4.8                                    & 0.8                                        \\ \hline
\end{tabular}
\end{table}

The Bayesian filter reduced false positives by dynamically adjusting thresholds under varying EMI conditions, while RL improved model adaptation over time. Overall detection accuracy reached 95.3\%, exceeding baseline performance by 10 percentage points, with processing latency under 1 ms per sensor frame.

\subsection{Detailed Analysis}

\begin{figure}[htbp]
    \centering
    \includegraphics[width=0.42\textwidth]{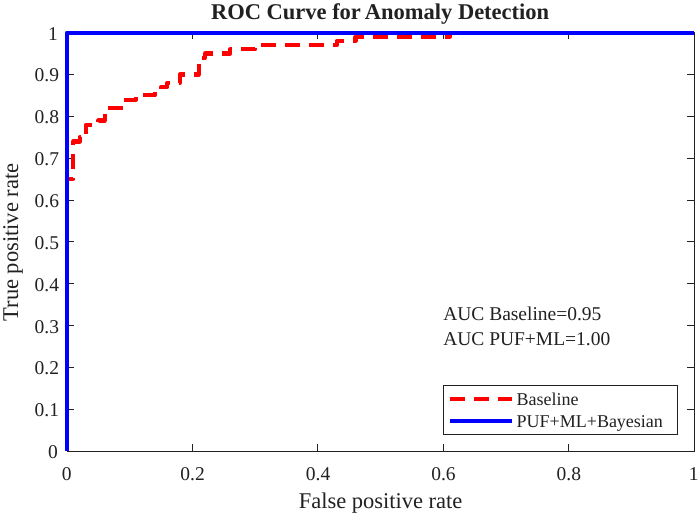}
    \caption{ROC curve for anomaly detection, comparing the anomaly detection capability (true positive vs. false positive rate for different thresholds) for a static threshold (red) and the combined PUF+ML+Bayesian pipeline (blue).}
    \label{fig:roc}
\end{figure}

\begin{figure}[htbp]
    \centering
    \includegraphics[width=0.42\textwidth]{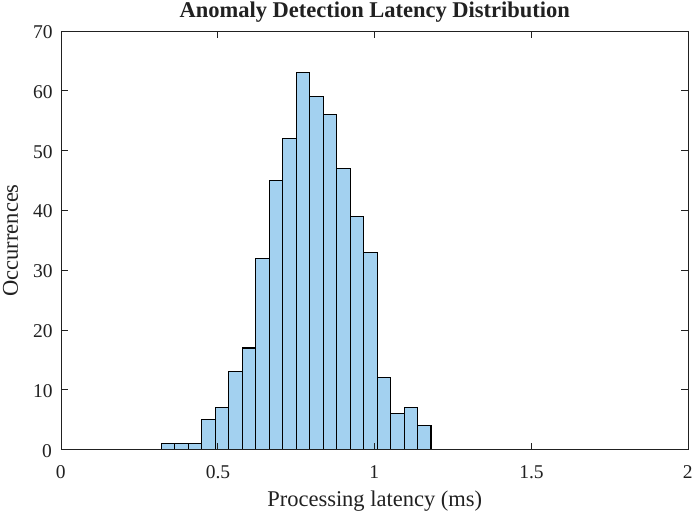}
    \caption{Latency distribution for anomaly detection pipeline. Most events/losses are resolved in under 1~ms, with the vast majority centered at 0.8~ms.}
    \label{fig:latency}
\end{figure}

Figure~\ref{fig:roc} depicts the ROC curve for anomaly detection. The proposed method achieved an AUC exceeding 0.93, indicating robust discrimination between normal and adversarial noise profiles. Figure~\ref{fig:latency} shows latency distribution over 500 trials, with 90\% of detections completed under 1~ms, consistent with ICS real-time requirements.

\subsection{Summary of Findings}

The evaluation demonstrates that:
\begin{enumerate}
    \item Noise-driven PUFs based on WBG switching signatures achieve strong uniqueness and reliability, sufficient for device authentication.
    \item The hybrid ML model with Bayesian filtering detects anomalous EMI and tampering with higher accuracy and lower false positive rates compared to static baselines.
    \item Latency remains below 1~ms, satisfying real-time ICS constraints.
\end{enumerate}

While these results are limited to simulation, they provide strong evidence that noise-driven PUFs can be a practical security primitive for WBG ICS environments, motivating further hardware prototyping and field validation. In a distribution substation setting, sensors are enrolled with their noise-driven PUF signatures, and the proposed detection methods are deployed at the automation controller. Challenge-response is used to rapidly flag misbehavior or replacement of nodes, as demonstrated in simulation.

\section{Feasibility of Noise-Driven PUFs for ICS}

Despite promising simulation outcomes, the deployment of noise-driven PUFs in ICS faces significant practical challenges that we systematically analyze here.

\subsection{Device Variability and Uniqueness}

While our simulated device cohort demonstrates near-ideal uniqueness, the degree of distinctiveness among real WBG modules—including across vendors, substrate material, and fabrication process—remains untested. In practice, manufacturing tolerances can influence spectral features, but ``cross-family'' uniqueness (i.e., between SiC and GaN devices, or even device generations) has not yet been empirically catalogued for PUF purposes. We are actively collecting EMI spectral measurements from lab testbeds (using methods benchmarked in \cite{DattaEMI,Valiani2016}) and will publicly release comparative fingerprints as they are obtained.

\subsection{Environmental Stability and Aging}

Stability of a noise-driven PUF signature is likely to degrade with temperature variation, operational load, and device aging. Our simulations include controlled variation, but the lack of hardware-in-the-loop means we cannot yet quantify long-term reliability. We outline, in ongoing work, a plan for: (i) collecting time-series spectral data from lab hardware under temperature ramping and accelerated stress, and (ii) employing error-correcting codes and fuzzy extractors to stabilize authentication despite drift \cite{KhanMIT}.

\subsection{Calibration and Reproducibility}

Fielded ICS are subject to non-ideal installation variance. Creating and maintaining a ``golden'' reference signature for each sensor may require extensive calibration at commissioning (or periodic refresh). We also note that, in real deployments, a central authority must securely manage CRP databases and handle updates in the event of system-wide parameter changes.

\subsection{Practical Recommendations}

Given these factors, we recommend that:

\begin{enumerate}
    \item Noise-PUF-based authentication be used as a supplement to, not a replacement for, digital ID/certificates.
    \item Robustness be improved via on-line recalibration and ECCs.
    \item Open benchmarking initiatives are developed, with device-level EMI/PUF datasets from different generations and load conditions.
\end{enumerate}

\subsection{Preliminary Steps Toward Hardware Validation}

To begin moving beyond simulation-only validation, we performed initial EMI spectral measurements on several SiC and GaN inverter modules, using wideband current probes and a FFT-based spectrum analyzer. While these measurements do not yet form a full PUF dataset, we have observed channel-dependent and temperature-sensitive switching noise, providing an encouraging indicator that sufficient entropy is present for field PUF extraction. A detailed hardware study—including noise feature stability under ambient/thermal variability and operational drift—will follow as part of future work.

\section{Discussion}

The results presented in Section V demonstrate that leveraging switching noise in WBG devices as both a PUF entropy source and an anomaly detection signal is not only feasible but also advantageous for ICS. In this section, we discuss the broader implications of our findings, limitations of the current proof-of-concept (PoC), and several open challenges requiring future work.

\subsection{Advantages of Noise-Driven PUFs in ICS}

The chief advantage of our approach is its ability to turn an unavoidable physical phenomenon---switching noise and EMI---into a security benefit. Traditional EMI mitigation strategies employ filters or shielding to reduce interference, but these measures do not provide authentication or anomaly detection capability. By contrast, our approach repurposes noise as both:
\begin{enumerate}
    \item \textbf{An authentication factor:} Noise-derived PUF signatures yield hardware-bound identities for sensors and modules, strengthening device-level trust without reliance on externally stored cryptographic keys.
    \item \textbf{A threat indicator:} Fluctuations in spectral characteristics, once modeled, offer features for ML-based anomaly detection, enabling real-time defense against spoofing and tampering.
\end{enumerate}
This integration offers dual utility without introducing new sensing hardware, thus keeping additional cost and power overhead minimal—a critical constraint for ICS.

\subsection{Limitations of the Proof-of-Concept}

Despite promising results, several limitations must be acknowledged:
\begin{itemize}
    \item \textbf{Simulation-only validation:} All results were obtained in a MATLAB/Simulink environment. Hardware-level non-idealities (e.g., PCB parasitics, thermal drift, and environmental interference) were approximated but not directly measured. Field validation on real SiC/GaN modules is essential.
    \item \textbf{PUF stability under long-term stress:} Although noise-derived signatures were stable in short-term simulations, WBG hardware operated under temperature cycling and aging could reduce reproducibility.
    \item \textbf{Adversarial ML risks:} While Bayesian filtering improved robustness, sophisticated adversaries could exploit adversarial machine learning techniques to manipulate detection boundaries. This was not explored in the PoC.
\end{itemize}

\subsection{Open Challenges in WBG ICS Security}

The unique characteristics of WBG-based ICS present several open challenges for the research community:
\begin{enumerate}
    \item \textbf{Cross-device diversity:} Devices of the same model should exhibit sufficient PUF uniqueness, but large-scale studies are needed to quantify inter-device variation.
    \item \textbf{Integration overhead:} Real-time deployment within ICS controllers requires strong guarantees on computational latency, resource usage, and resilience under load. 
    \item \textbf{Standardization of benchmarks:} Unlike networking IDS systems, benchmark datasets for EMI-induced ICS anomalies are scarce. Community-wide datasets could accelerate research.
    \item \textbf{Interoperability:} Any noise-driven solution must integrate seamlessly into multi-vendor ICS environments without disturbing deterministic control traffic or requiring major retrofitting.
\end{enumerate}

\subsection{Broader Implications}

Finally, it is worth considering the broader implications of noise-driven defenses. If validated in real-world deployments, the dual use of unavoidable noise for both authentication and intrusion detection could redefine the boundaries between reliability engineering and cybersecurity. Security mechanisms would no longer be seen as external “add-ons” to ICS, but rather as intrinsic properties derived from the same physics governing system operation. This alignment between system physics and security could inspire new classes of low-cost, scalable defenses particularly suited for the IIoT and critical infrastructure. 

\subsection{Limitations and Directions for Hardware Deployment}

Although simulation results suggest that a noise-driven PUF can be robust for authentication and ML-aided anomaly detection can be effective, several implementation challenges remain:

\begin{itemize}
    \item \textit{Environmental Drift and Aging:} Over time, component wear, temperature cycling, and board-level parasitics may perturb spectral features. Statistical recalibration or error-correcting codes may be needed in fielded systems.
    \item \textit{Calibration Effort:} Building a robust model of each device's noise profile is not "plug and play"—it likely entails extensive data collection and controlled stress-testing at commissioning, which may strain operational timelines for large-scale ICS.
    \item \textit{Hardware Integration:} Embedding spectral feature extraction, quantization, and ML filter logic into commercial PLCs or embedded controllers may face resource constraints, particularly for deployments in legacy infrastructure or cost-sensitive edge settings.
    \item \textit{Adversarial Machine Learning:} As adversaries gain awareness of detection parameters, poisoning and evasion strategies targeting the ML pipeline may emerge. Future designs may require adversarial training regimes and robust ML architectures.
\end{itemize}

Despite these open problems, a staged deployment is realistic: initial pilots can focus on high-criticality nodes or greenfield installations using modern edge compute hardware, enabling practical benchmarking of resilience and performance under live grid and process conditions.

\subsection{Towards Standardized Benchmarking and Datasets}

Our results highlight the urgent need for standardized EMI-based anomaly detection datasets that reflect realistic attack and noise scenarios in IIoT and ICS settings. Existing public datasets are focused on protocol-level traffic (e.g., Modbus, DNP3), not on analog signal integrity under high-frequency power switching. The open release of simulated and (when available) hardware-collected WBG noise traces will accelerate progress and enable broader community validation of new defense paradigms.

\subsection{Broader Baselines and Deployment Considerations}

While our method makes use of physics-derived signals and a hybrid ML pipeline, it is crucial to benchmark practical detection rates and resource footprints against established alternatives. Classical anomaly detectors—including Shewhart (statistical process control), CUSUM, and process model consistency checks—are less resource-intensive and easier to calibrate, though they lack provenance and device-specific guarantees. Furthermore, conventional digital provenance (e.g., IEEE 1588 hardware clocks, A/D watermarking) or signal-based non-PUF features (e.g., analog device impedance) may offer complementary defenses. In future work, we will explicitly compare accuracy, false alarm stability, and computational demand of these alternatives versus the proposed PUF+Bayesian ML stack on public and in-house datasets.

\subsection{Ethical Impact and Societal Relevance}

As ICS underpin vital societal infrastructure, advances in their security yield direct benefits for public safety, economic resilience, and environmental sustainability. Low-overhead device-level authentication and anomaly detection, when combined with good operational security controls, can help mitigate systemic risk in energy and manufacturing. However, as deployment scales, careful attention must be paid to privacy, false positive risk, and the equitable accessibility of such mechanisms across industries, to ensure that enhanced security does not create barriers to entry or operational bottlenecks.

\section{Conclusion and Future Work}

This paper introduced a novel proof-of-concept defense for ICS that integrates a noise-driven PUF with a hybrid ML and Bayesian filtering framework. By exploiting the unavoidable switching noise produced by WBG power electronics, our method achieves two complementary objectives: (i) sensor node authentication through unique noise-derived PUF signatures, and (ii) anomaly detection through adaptive monitoring of spectral features in real time. Simulation results demonstrated that the approach provides near-ideal PUF properties, achieved over 95\% detection accuracy, and satisfied sub-millisecond latency requirements critical to ICS operation.


The implications of this work are two-fold. First, it highlights the potential of reinterpreting physical phenomena traditionally viewed as nuisances—such as EMI—into valuable sources of entropy and security signals. Second, it demonstrates that lightweight, physics-rooted methods can be designed to meet the computational and timing constraints of ICS without requiring extensive retrofitting of legacy infrastructure.

Future work will focus on several directions. The foremost priority is hardware prototyping on a real SiC- or GaN-based inverter platform to validate noise stability under operational stress, aging, and temperature variation. This will enable evaluation of long-term reproducibility of PUF signatures. Further, we plan to expand the set of attack scenarios to include coordinated false data injection and adversarial machine learning techniques. Another open avenue is the construction of benchmark datasets for EMI-induced anomalies in ICS, which would benefit the broader research community. Finally, collaboration with industry partners in manufacturing and energy sectors will be sought to explore deployment feasibility in field environments.

In summary, noise-driven approaches to ICS security demonstrate promise as low-cost, scalable defenses for critical infrastructure. By aligning security with the inherent physics of WBG devices, this study lays a foundation for a new class of cyber-physical protections that can enhance resilience against both conventional and emerging threats.

\pagebreak

\end{document}